\documentclass[12pt]{article}
\usepackage{epsfig,amsfonts,amsmath,array,mathrsfs}

\newcommand{\pbs}[1]{\let\temp=\\#1\let\\=\temp}
\renewcommand{\theequation}{\thesection.\arabic{equation}}
\def\be{\begin{equation}}\def\ee{\end{equation}}
\def\cvp{\raise 2pt\hbox{,}} 
 \def\tr{\mathop{\rm tr}\nolimits}

 \def\d{{\rm d}}\def\nn{{\cal
N}}

 \def\Nf{N_{\mathrm f}}
 \def\uN{{\rm U}(N)} 
  
\def\F{\mathscr F} 

\def\wdv{W_{\text{DV}}}
 \def\R{\mathscr R}
\def\u{\text{U}(1)}
\def\wt{W_{\rm tree}}

\def\La{\Lambda}

\def\plb#1#2#3{{\it Phys.\ Lett.\ }{\bf B #1} (#2) #3}
\def\npb#1#2#3{{\it Nucl.\ Phys.\ }{\bf B #1} (#2) #3}

\def\jhep#1#2#3{{\it J. High Energy Phys.\ }{\bf #1} (#2) #3}
\def\prd#1#2#3{{\it Phys.\ Rev.\ }{\bf D #1} (#2) #3}

\def\atmp#1#2#3{{\it Adv.\ Theor.\ Math.\ Phys.\ }{\bf #1} (#2) #3}

\def\pr#1#2#3{{\it Phys.\ Rep.\ }{\bf #1} (#2) #3}

\begin{document}
%
%
\pagestyle{empty}
{\parskip 0in

\hfill LPTENS-07/03

\hfill hep-th/0701220}

\vfill
\begin{center}
{\LARGE The Chiral Ring and the Periods of the Resolvent}

\vspace{0.4in}

Frank F{\scshape errari}
\\
\medskip
{\it Service de Physique Th\'eorique et Math\'ematique\\
Universit\'e Libre de Bruxelles and International Solvay Institutes\\
Campus de la Plaine, CP 231, B-1050 Bruxelles, Belgique
}\\
\smallskip
{\tt frank.ferrari@ulb.ac.be}
\end{center}
\vfill\noindent

The strongly coupled vacua of an $\nn=1$ supersymmetric gauge theory
can be described by imposing quantization conditions on the periods of
the gauge theory resolvent, or equivalently by imposing factorization
conditions on the associated $\nn=\nobreak 2$ Seiberg-Witten curve (the
so-called strong-coupling approach). We show that these conditions are
equivalent to the existence of certain relations in the chiral ring,
which themselves follow from the fact that the gauge group has a
finite rank. This provides a conceptually very simple explanation of
why and how the strongly coupled physics of $\nn=1$ theories,
including fractional instanton effects, chiral symmetry breaking and
confinement, can be derived from purely semi-classical calculations
involving instantons only.

\vfill

\medskip
%
\begin{flushleft}
\today
\end{flushleft}
\newpage\pagestyle{plain}
\baselineskip 16pt
\setcounter{footnote}{0}

%
\section{Introduction}
\setcounter{equation}{0}

When a four dimensional $\nn=2$ supersymmetric gauge theory is
deformed into an $\nn=1$ theory, many interesting strong coupling
effects are expected to occur, like confinement, chiral symmetry
breaking and the creation of a mass gap. Whereas the solution of the
parent $\nn=2$ theory is governed by semi-classical instanton effects
\cite{SW}, most vacua of the corresponding $\nn=1$ theory are strongly
coupled and cannot be described in semi-classical terms. For example,
chiral observables vacuum expectation values are typically given by
fractional instanton series (that is to say, series for which the
expansion parameter is a fractional power of the usual instanton
factor).

Yet, it has been known for a long time that a very simple and
consistent description of the strongly coupled $\nn=1$ vacua could be
given in terms of the underlying $\nn=2$ theory \cite{SW}. This
so-called ``strong coupling'' approach is based on the fact that at
low energy, the parent $\nn=2$ theory is governed by a free abelian
gauge theory. \emph{Assuming} that the $\nn=1$ theory creates a mass
gap, the $\nn=2$ moduli must be frozen at the singularities of the
$\nn=2$ moduli space when the $\nn=1$ deformation is turned on. This
is so because the only way a free abelian gauge theory can have a mass
gap is through the usual Higgs mechanism, and this mechanism can only
occur when charged fields, that are only present at the singularities,
condense. This strong coupling procedure then implies confinement and
chiral symmetry breaking \cite{SW}, and, when combined with the
generalized Konishi anomaly equations \cite{CDSW}, essentially fixes
all the correlators of chiral operators in the theory.

A second, \emph{equivalent} description of $\nn=1$ theories can be
given in the context of the gauge theory/matrix model correspondence
\cite{DV,PR}. This is an elegant approach that allows to derive most
of the known exact results in the field, including the Seiberg-Witten
solution of $\nn=2$ super Yang-Mills \cite{SW,CV}, and many
non-perturbative effects on the space of vacua of $\nn=1$ theories
\cite{fer1,fer2,phase1,phase2}. On the matrix model side of the
correspondence, the most general solution depends on a set of
parameters, the filling fractions, that can be chosen arbitrarily. On
the gauge theory side, however, these filling fractions correspond to
the gluino condensates $S_{I}$, and thus must be fixed,
non-perturbative functions of the parameters (couplings in the
tree-level superpotential and dynamically generated scale or gauge
coupling constant).

Understanding the basic principles that fix the filling fractions
$S_{I}$ in the gauge theory is a major challenge.
The original conjecture was that the filling fractions are determined
by extremizing a certain superpotential, the so-called Dijkgraaf-Vafa
glueball superpotential $\wdv(S_{I})$,
\be\label{wdveq}\frac{\partial\wdv}{\partial S_{I}} = 0\, .\ee
This proposal is well motivated by using the gauge/string
correspondence (in the dual string formulation, the Dijkgraaf-Vafa
superpotential is a flux superpotential \cite{CIV}), but is difficult
to understand from the field theory point of view. The equations
\eqref{wdveq} look like complicated dynamical constraints,
consistently with the idea that a field theoretic proof would involve
a deep understanding of the non-perturbative gauge dynamics
\cite{CDSW}. Actually, we are going to show that a conceptually simple
justification can be found.

\begin{figure}
\centerline{\epsfig{file=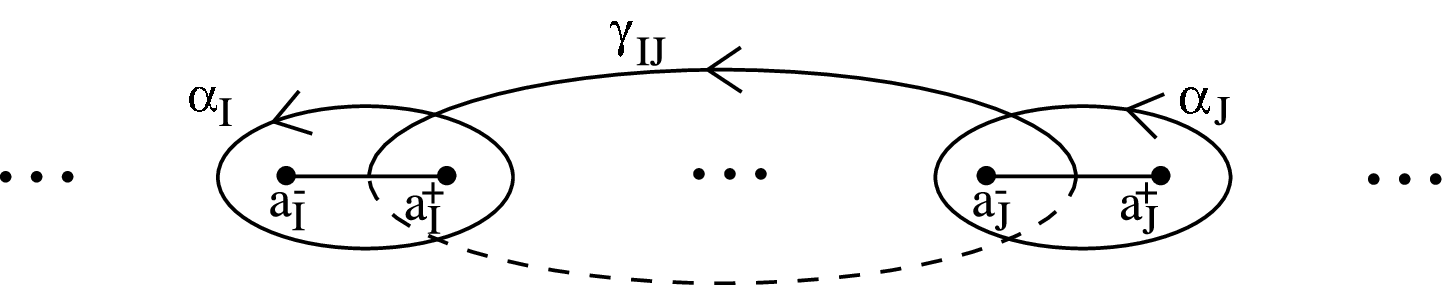,width=14cm}}
\caption{The non-compact two-sheeted Riemann surface $\mathcal C$,
with the contours $\alpha_{I}$ and $\gamma_{IJ}$ used in
the main text.
\label{fig1}}
\end{figure}

A very nice property of the equations \eqref{wdveq}, pointed out in
\cite{phase2}, is that they are mathematically equivalent to a set of
quantization conditions for the periods of the one-form $\R\d z$,
where $\R$ is the gauge theory resolvent defined by
\be\label{resdef}
\R(z) = \tr\frac{1}{z-X}\, \cdotp\ee
The adjoint chiral superfield $X$ in the above formula is the
superpartner of the vector superfield in the $\nn=2$ theory. A short
proof of this statement is given for example in \cite{fer0}. More
precisely, the generalized Konishi anomaly equations \cite{CDSW} imply
that $\R\d z$ is a meromorphic differential on a certain hyperelliptic
curve $\mathcal C$. If we write down the equation for $\mathcal C$ in
the form
\be\label{RS} \mathcal C:\ y^{2} = 
\prod_{I=1}^{\tilde d}(z-a_{I}^{-})(z-a_{I}^{+})\, ,\ee
and define the contours as in Figure \ref{fig1}, the quantization
conditions read\footnote{The conditions \eqref{qc1} are equivalent to
$\partial\wdv/\partial S_{I}=\partial\wdv/\partial S_{J}$. As
explained in the Section 4 of \cite{fer0}, the missing equation can
then be derived from a standard Ward identity.}
\begin{align}\label{qc1}
\oint_{\gamma_{IJ}}\!\R\,\d z\in 2i\pi\mathbb Z\, ,\\
\label{qc2}\oint_{\alpha_{I}}\!\R\,\d z\in 2i\pi\mathbb Z\, .
\end{align}
The quantization conditions \eqref{qc2} might seem obvious, because
$\frac{1}{2i\pi}\oint_{\alpha_{I}}\!\R\,\d z$ may be interpreted as
giving the number of eigenvalues of the matrix $X$ in the cut
$[a_{I}^{-},a_{I}^{+}]$. However, as we explain in the next Section,
\eqref{qc2} is non-trivial from a fully non-perturbative point of
view. Actually, both the quantization conditions \eqref{qc1} and
\eqref{qc2} appear on an equal footing in the arguments that we
present in this paper.

So we have two simple, elegant and physically well-motivated
procedures to derive the solution of $\nn=1$ gauge theories that are
deformations of parent $\nn=2$ gauge theories. It is known that these
procedures are mathematically equivalent. Naively, it is a priori very
difficult to justify these approaches from first principles. This
might seem inevitable, since they are at the basis of the derivation
of strongly coupled effects that cannot be described in semi-classical
terms.

The main result of the present work is to provide an extremely simple
argument, from first principles, proving directly the quantization
conditions \eqref{qc1} and \eqref{qc2}, or equivalently the validity
of the strong coupling approach. The main idea of the proof is to
concentrate on relations in the chiral ring that must exist because
the gauge group is of finite rank. The existence of these relations is
a basic difference with the associated matrix model, for which the
size of the matrix is infinite. The main point is that the conditions
\eqref{qc1} and \eqref{qc2} are simply equivalent to a particular form
of the constraints. Since the constraints are operator relations, they
must remain true in all the vacua of the deformed $\nn=1$ theory if
they are established in the $\nn=2$ limit. This is possible if and
only if \eqref{qc1} and \eqref{qc2} are satisfied in all the $\nn=1$
vacua, or equivalently if and only if the $\nn=1$ vacua are described
by the usual factorized Seiberg-Witten curves.

We give full details in the case of the $\nn=2$ $\uN$ theory deformed
by a superpotential term $\tr\wt(X)$, including when $\Nf\leq 2N$
flavors of quarks are present. However, our arguments are completely
general. The same ideas actually apply as well to $\nn=1$ theories
that are not necessarily deformations of $\nn=2$ theories.

The paper is organized as follows. In Section 2, we discuss the
constraints in the chiral ring that follow from the finiteness of the
number of colors $N$. These constraints are at the basis of the main
argument that is explained in Section 3. Finally, in Section 4 we
present some open problems and discuss the relations of the present
work with \cite{fer0}.

\noindent\textbf{A note on notations}: in the following, we consider
expectation values of chiral operators $\mathscr O$ in various vacua.
The symbol $\langle\mathscr O\rangle$ is used when we do not need to
specify a particular vacuum, typically in expressions that are valid
in all the vacua. Relations valid in all the vacua are also often
noted as operator relations\footnote{By operator relations, we always
mean operator relations in the chiral ring.} without brackets. On the
other hand, in vacuum-dependent equations, we always specify
explicitly in which vacuum $|0\rangle$ we are working, by using the
symbol $\langle 0|\mathscr O|0\rangle$.

\section{On the relations in the chiral ring}
\setcounter{equation}{0}

Let us consider the $\uN$ theory with $\nn=1$ supersymmetry and one
adjoint chiral superfield $X$. The generalization to the theory with
flavors is discussed in 3.4.

The chiral ring of the theory is generated by the operators \cite{CDSW}
\be\label{defgen} u_{k}=\tr X^{k}\, ,\quad u_{k}^{\alpha} =
\frac{1}{4\pi}\tr W^{\alpha}X^{k}\, ,\quad v_{k} =
-\frac{1}{16\pi^{2}}\tr W^{\alpha}W_{\alpha} X^{k}\, ,\ee
where $W^{\alpha}$ is the $\nn=1$ super field strength. It is
well-known that for any $N\times N$ matrix $X$, the traces $\tr X^{k}$
for $k>N$ can be expressed in terms of the $\tr X^{k}$ for $1\leq
k\leq N$. These relations take the form
\be\label{cleq} u_{N+p} = \mathscr
P_{\text{cl},\,p}(u_{1},\ldots,u_{N})\, ,\quad p\geq 1\, ,\ee
where the $\mathscr P_{\text{cl},\,p}$ are homogeneous polynomials of
degree $p+N$ in the $u_{1},\ldots,u_{N}$ ($u_{k}$ being of degree
$k$). Let us emphasize that these relations are simply
\emph{identities,} that follow from the finiteness of the rank of the
gauge group. There are also similar relations relating the
$u^{\alpha}_{N+p}$ and the $v_{N+p}$ for $p\geq 0$ to the
$u^{\alpha}_{0},\ldots,u^{\alpha}_{N-1},u_{1},\ldots,u_{N}$ and
$v_{0},\ldots,v_{N-1},u_{1},\ldots,u_{N}$ respectively, but we don't
need them for our analysis.

Non-perturbatively, the relations \eqref{cleq} may be modified. The
new relations must be consistent with the $\u_{\text A}$ and
$\u_{\text R}$ symmetries of the theory. These symmetries act on the
operators, the parameters $g_{k}$ (defined in \eqref{wtsp1}) and the
instanton factor $\La^{2N}$ as
\be\label{asign1}
\begin{matrix}
& u_{k} & u^{\alpha}_{k} & v_{k} & \La^{2N} & 
g_{k} \\
{\rm U}(1)_{\rm A} & k & k & k & 2N & -k-1 \\
{\rm U}(1)_{\rm R} & 0 & 1 & 2 & 0 &\hphantom{,\,} 2 \, .
\end{matrix}\ee
The $\u_{\text R}$ symmetry implies that the $u_{k}$ cannot mix with
the other operators $u^{\alpha}_{k}$ or $v_{k}$, and that the quantum
version of the relations \eqref{cleq} cannot depend on the couplings
$g_{k}$. In other words, we can restrict ourselves (and this will turn
out to be sufficient for our purposes) to the sector of the chiral
ring with zero $\u_{\text R}$ charge. This sector is itself a ring
$A$, generated by the $u_{k}$. The relations in $A$ can take the
general form
\be\label{qrel} u_{N+p} = \mathscr
P_{p}(u_{1},\ldots,u_{N};\La^{2N})\, ,\quad p\geq 1\, ,\ee
where the $\mathscr P_{p}$ are polynomials of $\u_{\text A}$ charge
$p+N$ that go to the classical polynomials $\mathscr P_{\text{cl},\,
p}$ when $\La^{2N}$ goes to zero.\footnote{Note that only the
instanton factor $\La^{2N}$ can enter, by $2\pi$ periodicity in the
$\theta$ angle, because \eqref{qrel} is an operator relation and is
thus valid in all the vacua. Fractional instanton effects do occur in
$\nn=1$ theories, but only in relations that are valid in a particular
vacuum (or a particular set of vacua).}

A fact we would like to emphasize is that the quantum corrections in
\eqref{qrel} do not represent a ``deformation,'' in any sensible
mathematical sense, of the classical chiral ring. The classical chiral
ring $A_{\text{cl}}$ (in the sector of zero R-charge we're interested
in) is simply the polynomial algebra generated by the $u_{k}$ for
$1\leq k\leq N$. The quantum version $A$ of this ring must be
commutative (since only bosonic variables are present) and generated
freely by the same elements $u_{1},\ldots,u_{N}$. This implies that
\be\label{Arel} A=A_{\text{cl}}=\mathbb C[u_{1},\ldots,u_{N}]\, .\ee
A more abstract way to understand this is to note that because the
quantum theory can be seen as a smooth deformation of the classical
theory obtained by turning on the instanton factor $\La^{2N}$, the
possible deformations of $A_{\text{cl}}$ can be studied using the
standard deformation theory based on Hochschild cohomology. It is an
elementary result that there is no possible non-trivial deformation of
a polynomial algebra that preserves commutativity (see for example
\cite{deformation} for an elementary exposition). The conclusion is
that the zero R-charge sector of the chiral ring is \emph{not} quantum
corrected.\footnote{The full chiral ring can be deformed, because for
example elements that are nilpotent classically, like the glueball
operator, are not in the quantum theory \cite{CDSW}. However these
deformations are not related to the existence of the quantum-corrected
relations \eqref{qrel}.}

So what is the interpretation of the relations \eqref{qrel}? They
actually represent the \emph{definitions} of what we call the
$u_{N+p}$ for $p\geq 1$. These definitions are \emph{non-dynamical,}
and a priori can be completely arbitrary (as long as they are
consistent with symmetries and the classical limit). This freedom has
actually been used in some instances in the literature (for example to
match results obtained by different methods \cite{freevar,instrev}).
In particular, it is perfectly consistent to define the $u_{k}$ by the
``classical'' relations \eqref{cleq}, even in the full quantum theory.
However, depending on the context, a natural non-perturbative
definition of what is called $u_{N+p}$ for $p\geq 1$ may involve
quantum-corrected relations of the form \eqref{qrel}.

For the purpose of our investigations, the natural variables are such
that the anomaly equations, that were derived in perturbation theory
in \cite{CDSW}, take the same form in the full quantum theory: for all
$n\geq -1$,
\begin{align}\label{a1}
-N\sum_{k\geq 0}g_{k}u_{n+k+1}+2\sum_{q_{1}+q_{2}=n}\bigl(
v_{q_{1}}u_{q_{2}}+u^{\alpha}_{q_{1}}u_{q_{2}\,\alpha}\bigr)&=0\, ,\\
\label{a2}
-N\sum_{k\geq 0}g_{k}u^{\alpha}_{n+k+1}+2\sum_{q_{1}+q_{2}=n}
v_{q_{1}}u^{\alpha}_{q_{2}}&=0\, ,\\
\label{a3}
-N\sum_{k\geq 0}g_{k}v_{n+k+1}+\sum_{q_{1}+q_{2}=n}
v_{q_{1}}v_{q_{2}}&=0\, .
\end{align}
Note that this assumption does not mean that the equations do not get
non-per\-tur\-ba\-ti\-ve corrections, but rather that the
non-perturbative corrections can be absorbed in a proper definition of
the variables. We also expect that this definition of the variables is
the same as the one that enters naturally in the context of instanton
calculus \cite{instrev}.

Lacking a detailed non-perturbative analysis of the anomaly equations,
we shall allow the relations \eqref{qrel} to take the most general
possible form a priori. This implies an interesting subtlety that has
been overlooked in previous works. It is clear that the gauge theory
resolvent \eqref{resdef} does depend explicitly on the particular
definitions of the higher moments $u_{N+p}$. In particular, there are
many consistent definitions, with relations of the form \eqref{qrel},
that violate the quantization conditions \eqref{qc2}. A nice feature
of our approach is that we don't need to assume \eqref{qc2}, and both
quantization conditions \eqref{qc1} and \eqref{qc2} will be derived at
the same time.

\section{Relations and the quantization of the periods}
\subsection{Picking a suitable vacuum}

An important point is that the precise form of the constraints
\eqref{qrel} can be derived from a purely semi-classical analysis.
This is possible because we can always find a vacuum that is both
arbitrarily weakly coupled and suitable to fix the relations
\eqref{qrel} unambiguously.

For example, the R-symmetry implies that the polynomials $\mathscr
P_{p}$ in \eqref{qrel} cannot depend on the couplings in the
tree-level superpotential. We can thus choose the latter at our
convenience. We pick a degree $N+1$ superpotential such that
\be\label{wtsp1} \wt'(x) = \sum_{k=0}^{N} g_{k}
x^{k}=g_{N}\prod_{I=1}^{N}(x-a_{I})=g_{N}P_{N}(x)\, .\ee
Classically, the gauge theory has several vacua, depending on the
numbers $N_{I}\geq 0$ of eigenvalues of the adjoint field $X$ that are
taken to be equal to $a_{I}$. These classical vacua are denoted by
$|N_{1},\ldots,N_{N}\rangle$ and correspond to a pattern of gauge
symmetry breaking $\uN\rightarrow\text U(N_{1})\times\cdots\times\text
U(N_{N})$. Let us focus on the Coulomb vacuum
\be\label{Cvac}|\text C\rangle = |1,\ldots,1\rangle\, ,\ee
in which the low energy gauge group is $\u^{N}$. This vacuum can be
made arbitrarily weakly coupled by going to the region
$|a_{I}-a_{J}|\gg |\La|$ in the space of parameters. Moreover, in this
vacuum, the expectation values $\langle\text C| u_{1}|\text
C\rangle,\ldots,\langle\text C| u_{N}|\text C\rangle$ are independent,
unconstrained variables.\footnote{This is not true in general. For
example, for a vacuum $|0\rangle$ with an unbroken gauge group and for
which all the eigenvalues of $X$ are equal classically, we have
automatically $\langle 0| u_{k}|0\rangle =N^{1-k}\langle 0| u_{1}|0
\rangle^{k}$ at the perturbative level. This shows that there is only
one independent variable. In the non-perturbative theory, the
relations are modified by fractional instanton effects, but the number
of independent variables do not change.} Equivalently, we can take the
$g_{N}\rightarrow 0$ limit of the $\nn=2$ theory, in which case the
$u_{1},\ldots,u_{N}$ are moduli. This means that if we can find
polynomials $\hat{\mathscr P_{p}}$ such that
\be\label{qrelvacC} \langle\text C| u_{N+p}|\text C\rangle =
\hat{\mathscr P}_{p}\bigl(\langle\text C| u_{1}|\text
C\rangle,\ldots,\langle\text C| u_{N}|\text C\rangle;\La^{2N}\bigr)\,
,\quad p\geq 1\, ,\ee
for arbitrary values of the $\langle\text C| u_{k}|\text C\rangle$,
then we know automatically that
\be\label{fundPP}\mathscr P_{p} = \hat{\mathscr P}_{p}\, .\ee
The above reasoning is quite powerful: we are able to derive operator
relations by studying the theory in a particular vacuum. This is a
basic feature of our method. It is made possible by the fact that we
know \emph{a priori} that operator equations of the form \eqref{qrel}
must exist.

\subsection{Example}

Let us use the above idea to compute the polynomials $\mathscr P_{p}$
in the theory with no flavor. Equations \eqref{a1} and \eqref{a3} can
be easily solved (using in particular $\langle u^{\alpha}_{k}\rangle =
0$ by Lorentz invariance) to yield a general formula for the gauge
theory resolvent expectation value, valid in any vacuum \cite{CDSW}.
For the degree $N+1$ tree level superpotential \eqref{wtsp1}, and
introducing degree $N-1$ polynomials $Q_{N-1}$ and $R_{N-1}$ (whose
precise forms depend on the particular vacuum under consideration), we
have
\be\label{Rgen} \bigl\langle\R(z)\bigr\rangle =
\frac{Q_{N-1}(z)}{\sqrt{P_{N}(z)^{2} - R_{N-1}(z)}}\,\cdotp
\ee
Note that since
\be\label{Rasy}\bigl\langle\R(z)\bigr\rangle
\underset{z\rightarrow\infty}{\sim}\frac{N}{z}\,\cvp\ee
we know that $Q_{N-1}(z) = Nz^{N-1}+\cdots$ in all the vacua.

A useful property of the Coulomb vacuum for the theory with no flavor
is that the $\u_{\text A}$ symmetry \eqref{asign1} implies that the
$\langle\text C|u_{k}|\text C\rangle$ cannot get quantum corrections
for $k\leq 2N-1$. This is so because in the Coulomb vacuum the quantum
corrections are entirely generated by instantons, and the instanton
factor $\La^{2N}$ has $\u_{\text A}$ charge $2N$. We thus obtain
\be\label{ukC} \langle\text C| u_{k}|\text C\rangle =
\sum_{I=1}^{N}a_{I}^{k}\, ,\quad 1\leq k\leq 2N-1\, ,\ee
which is equivalent to the following asymptotic condition,
\be\label{Rasydetailed} \langle\text C|\R(z)|\text C\rangle
= \frac{P'_{N}(z)}{P_{N}(z)} + \mathcal O(1/z^{2N+1})\, .\ee
Plugging \eqref{Rgen} into \eqref{Rasydetailed}, multiplying by
$\sqrt{P_{N}^{2}-R_{N-1}}$ and expanding at large $z$ immediately
yield
\be\label{inter1} Q_{N-1}(z) = P'_{N}(z) + \mathcal O(1/z^{2})\, .\ee
Since both $Q_{N-1}$ and $P'_{N}$ are polynomials, we must have
\be\label{res1} Q_{N-1} = P'_{N}\, .\ee
Taking this result into account, \eqref{Rasydetailed} implies that
\be\label{inter2}\frac{1}{\sqrt{P_{N}(z)^{2}-R_{N-1}(z)}} =
\frac{1}{P_{N}(z)} + \mathcal O(1/z^{3N})\, .\ee
Inverting this relation, taking the square and expanding at large $z$
then yields
\be\label{inter3} P_{N}(z)^{2}-R_{N-1}(z) =  P_{N}(z)^{2} + \mathcal
O(1)\, ,\ee
or equivalently that $R_{N-1}$ must be a constant $r$,
\be\label{RNeq} R_{N-1}(z) = r\, .\ee
The relation \eqref{Rasydetailed}, and thus \eqref{res1} and
\eqref{RNeq}, are of course valid only in the Coulomb vacuum.

We have thus achieved our goal: all the polynomials $\mathscr P_{p}$
in \eqref{qrel} can be expressed in terms of the constant $r$, by
expanding $\langle\text C|\R(z)|\text C\rangle$ at large $z$. For
example, the first non-trivial correction is obtained for $\mathscr
P_{N}$ and reads

\be\label{Pex1}\mathscr P_{N} = \mathscr P_{\text{cl},\, N} + \frac{N 
r}{2}\,\cdotp\ee
Clearly $r$ must be proportional to $\La^{2N}$ and, at the expense of
rescaling $\La$, we can always choose $r=4\La^{2N}$, which is the
standard convention.\footnote{The identification of the constant $r$
with $4\La^{2N}$ is straightforward in the present case, but it plays
no r\^ole in the following, and in particular is not needed to prove
the quantization conditions \eqref{qc1} and \eqref{qc2}.} 

\subsection{Encoding the form of the relations}

In the theory with no flavor, we have been able to compute the
polynomials $\mathscr P_{p}$ by using a simple symmetry argument. This
would not be the case for more general theories, for example when a
large number of flavors are present. However, the only important point
for us is that it is always possible to do this calculation in a
purely semi-classical context (see also Section 4 for another possible
way to derive the relations).

This being said, let us now show that these relations are
\emph{equivalent} to a simple algebraic equation satisfied by the
quantum characteristic function
\be\label{defF}\F(z) = \det (z-X)\, .\ee
This algebraic relation follow from a simple procedure\footnote{This
trick has appeared several times in the literature, for example in
\cite{svrcek} and \cite{fer0}.} to compute the polynomials
$u_{N+p}=\mathscr P_{p}$ recursively. The characteristic function $\F$
admits a simple expansion in terms of the $u_{k}$ of the form
\be\label{Fexp} \F(z) = z^{N}-\sum_{k\geq 1}F_{k}z^{N-k}\, ,\ee
where the $F_{k}=u_{k}/k+\cdots$ are polynomials in the $u_{q}$s of
$\u_{\text A}$ charge $k$. The $F_{k}$s can be computed explicitly by 
writing
\be\label{exprep} \F(z) = \det (z-X) =z^{N}e^{\tr\ln (1-X/z)}
= z^{N}e^{-\sum_{k\geq 1}u_{k}/(kz^{k})}\ee
and expanding at large $z$. At the classical level, $\F$ is a
polynomial of degree $N$. We thus have $F_{k}=0$ for all $k>N$. Since
$F_{k}$ is the sum of $u_{k}/k$ plus terms that depend only on the
$u_{q}$ for $q<k$, this yields convenient recursion relations that
determine all the polynomials $\mathscr P_{\text{cl},\, p}$. Quantum
mechanically, we can find $\langle\F\rangle$ in the Coulomb vacuum
from the formula
\be\label{RCoulomb} \langle\text C|\R(z)|\text C\rangle =
\frac{P'_{N}(z)}{\sqrt{P_{N}(z)^{2} - 4\La^{2N}}} \ee
derived in the previous subsection, by integrating the relation
\be\label{FRrel} \frac{\d}{\d z}\ln\bigl\langle\F(z)\bigr\rangle =
\bigl\langle\R(z)\bigr\rangle\ee
and using
\be\label{Fasy}\bigl\langle\F(z)\bigr\rangle
\underset{z\rightarrow\infty}{\sim}z^{N}\, .\ee
This yields
\be\label{Fform}\langle\text C|\F(z)|\text C\rangle =
\frac{1}{2}\Bigl( P_{N}(z) + \sqrt{P_{N}(z)^{2} - 4\La^{2N}}\Bigr)\,
.\ee
In particular, $\langle\text C|\F(z)|\text C\rangle$ is not a
polynomial anymore, but it satisfies a simple quadratic equation
\be\label{eqF} \langle\text C|\F(z)|\text C\rangle +
\frac{\La^{2N}}{\langle\text C|\F(z)|\text C\rangle} = P_{N}(z)\, .\ee
By expanding at large $z$, this equation yields the $\langle\text
C|u_{k}|\text C\rangle$ for $k>N$ as a function of the $\langle\text
C| u_{1}|\text C\rangle,\ldots,\langle\text C| u_{N}|\text C\rangle$.
To see how this works in details, let us write
\be\label{FFexp}\bigl\langle\F(z)\bigr\rangle = z^{N} - \sum_{k\geq
1}\bigl\langle F_{k}\bigr\rangle z^{N-k}\, ,\quad
\frac{1}{\bigl\langle\F(z)\bigr\rangle} = z^{-N} + \sum_{k\geq
1}\bigl\langle \tilde F_{k}\bigr\rangle z^{-N-k}\, ,\ee
and plug these expansions into \eqref{eqF}. Since the right hand side
is a polynomial, all the terms with negative powers of $z$ must cancel
in the left hand side, yielding
\be\label{recforF} \langle\text C| F_{p+N}|\text C\rangle = \La^{2N}
\langle\text C| \tilde F_{p-N}|\text C\rangle\quad\text{for all}\
p\geq 1\, ,\ee
with the convention that $\tilde F_{0}=1$ and $\tilde F_{k}=0$ if
$k<0$. Since $\tilde F_{k} = u_{k}/k+\cdots$ depends only on the
$u_{q}$ for $q\leq k$, \eqref{recforF} generates recursively all the
relations \eqref{qrel}.

Now comes the main point of our argument. The equations
\eqref{recforF} not only provide a simple way to fix unambiguously the
relations \eqref{qrel},\footnote{Here we use the fact, already
emphasized in Section 3.1, that the variables $\langle\text C|
u_{1}|\text C\rangle,\ldots\langle\text C|u_{N}|\text C\rangle$ are
independent in the Coulomb vacuum.} they are actually
\emph{equivalent} to them. Since \eqref{qrel} is valid in all the
vacua, it must be so for \eqref{recforF}. In other words, \emph{we
have shown that the equations \eqref{recforF} are operator relations
valid in all vacua,}
\be\label{oprelFk} F_{p+N} = \La^{2N}\tilde F_{p-N}\, ,\quad p\geq 1\,
,\ee
because they simply correspond to a convenient rewriting of the
operator relations \eqref{qrel}. Moreover, since the couplings $g_{k}$
cannot appear in \eqref{qrel}, we know that \eqref{oprelFk} must be
true for any tree-level superpotentiel $\tr\wt(X)$, not necessarily of
the form \eqref{wtsp1}.

Using the operator relations \eqref{oprelFk}, we deduce that
$\langle\F(z)\rangle + \La^{2N}/\langle\F(z)\rangle$ has no negative
powers of $z$ in its large $N$ expansion, not only in the Coulomb
vacuum but also \emph{in all the other vacua} of the $\nn=1$ theory.
Equivalently, this implies that
\be\label{oprelF} \bigl\langle\F(z)\bigr\rangle +
\frac{\La^{2N}}{\bigl\langle\F(z)\bigr\rangle} = P(z)\ee
is a polynomial in all vacua. The precise form of $P(z)=z^{N}+\cdots$
depends on the particular vacuum because the operator relations
\eqref{recforF} do not constrain the positive powers in $z$ in the
left hand side of \eqref{oprelF}.

The fundamental point in our argument is that the algebraic equation
\eqref{oprelF} is not dynamical, but rather acts as a generating
equation for the relations \eqref{qrel}. It is very important that the
algebraic equation satisfied by $\F$ contains only this purely
``kinematical'' information. Again, this is why we can derive that the
equation must be valid in all the vacua of the gauge theory.

\subsection{The quantization conditions}

We now have all the necessary ingredients to show that the
quantization conditions \eqref{qc1} and \eqref{qc2} must always be
valid. First of all, from \eqref{FRrel} and \eqref{oprelF} we find
\begin{align}\label{Fgen2}\bigl\langle\F(z)\bigr\rangle
&=\frac{1}{2}\Bigl( P(z) + \sqrt{P(z)^{2} - 4\La^{2N}}\Bigr)\, ,\\
\label{Rgen2} \bigl\langle\R(z)\bigr\rangle &=
\frac{P'(z)}{\sqrt{P(z)^{2} - 4\La^{2N}}}\,\cdotp
\end{align}
Both $\langle\F\rangle$ and $\langle\R\rangle$ are thus meromorphic
functions defined on the \emph{same} Riemann surface $\mathcal C$,
independently of the vacuum under consideration. The equations
\eqref{Fgen2} and \eqref{Rgen2} have a form similar to \eqref{Fform}
and \eqref{RCoulomb}, but the degree $N$ polynomial $P$ is
vacuum-dependent and is not equal to $P_{N}$ in general. In particular
$\wt'$ can be any polynomial, not necessarily of the form
\eqref{wtsp1}. Consistency with the anomaly equations \eqref{a1} and
\eqref{a3} for arbitrary $\wt'$ of degree $d$ actually immediately
implies the existence degrees $d-1$, $N-\tilde d$ and $d-\tilde d$
polynomials $\Delta_{d-1}$, $F_{N-\tilde d}$ and $H_{d-\tilde d}$
respectively such that
\be\label{fact} \wt'(z)^{2} - \Delta(z) = H_{d-\tilde d}(z)^{2}y^{2}\,
,\quad P(z)^{2}-4\La^{2N} = F_{N-\tilde d}(z)^{2}y^{2}\, .\ee
These are the factorization equations \eqref{fact} at the basis of the
strong coupling approach to $\nn=1$ gauge theories! In particular, the
curve
\be\label{SWcurve} Y^{2} = P(z)^{2} - 4\La^{2N}\ee
is the Seiberg-Witten curve of the $\nn=2$ theory obtained in the
$\wt\rightarrow 0$ limit.

The fact that $\F$ and $\R$ are defined on the same Riemann surface is
also all we need to derive the quantization conditions \eqref{qc1} and
\eqref{qc2}. Indeed, in general, integrating \eqref{FRrel} taking into
account \eqref{Fasy} yields
\be\label{intforF}\bigl\langle\F(z)\bigr\rangle =
\lim_{\mu_{0}\rightarrow\infty}\Bigl(\mu_{0}^{N}\exp\int_{\mu_{0}}^{z}
\bigl\langle\R(z')\bigr\rangle\d z'\Bigr)\, ,\ee
where $\mu_{0}$ is a point on the first sheet (the sheet for which
\eqref{Fasy} is valid) of the Riemann surface. This formula shows that
$\langle\F(z)\rangle$ is generically a multivalued function on the
curve $\mathcal C$ on which $\langle\R(z)\rangle$ is well-defined,
because one must specify the contour from the point at infinity
$\mu_{0}$ to $z$ to do the integral in \eqref{intforF}. The integral
representation also shows that $\langle\F(z)\rangle$ will be single
valued \emph{if and only if} the quantization conditions
\be\label{qc} \oint_{\gamma_{IJ}}\!\bigl\langle\R\bigr\rangle\,\d z\in
2i\pi\mathbb Z\, ,\quad
\oint_{\alpha_{I}}\!\bigl\langle\R\bigr\rangle\,\d z\in 2i\pi\mathbb
Z\, , \ee
are satisfied. We have thus completed the proof of \eqref{qc1} and
\eqref{qc2}.

\subsection{Generalization to the case with flavors}

Let us now add $\Nf= 2N$ flavors of fundamental and antifundamental
quarks and antiquarks $Q_{q}$ and $\tilde Q^{q}$. The other cases
$\Nf<2N$ can be obtained by integrating out some of the flavors. The
tree-level superpotential has the form
\be\label{wtfla}W = \tr\wt(X) + \sum_{q =1}^{2N}{}^{T}\tilde
Q^{q}(X-m_{q})Q_{q}\, .\ee
The most general classical vacuum $|N_{I};\nu_{q}\rangle$ is specified
by the numbers of eigenvalues of the matrix $X$, $N_{I}\geq 0$ and
$\nu_{Q}=0$ or 1, that are equal to the $I^{\text{th}}$ extremum of
$\wt$ and $m_{q}$ respectively \cite{phase2}. In particular,
\be\label{sumeigen}\sum_{I}N_{I} + \sum_{q}\nu_{q} = N\, .\ee
The pattern of gauge symmetry breaking in a vacuum $|N_{I};\nu_{q}\rangle$
is $\uN\rightarrow\prod_{I}\text U(N_{I})$.

The anomaly equations \cite{seifla} and associated matrix model
\cite{addfla} for this theory are well-known, and we shall not repeat
the details here (a very detailed discussion is included in
\cite{PR}). The chiral ring relations are still of the general form
\eqref{qrel}, but the instanton factor is now
\be\label{ifac} h = e^{2i\pi\tau}\ee
and the polynomials $\mathscr P_{p}$ can depend symmetrically on the
masses $m_{q}$ (but cannot depend on the couplings in $\wt$).

We can compute the $\mathscr P_{p}$ semi-classically by choosing $\wt$
of the form \eqref{wtsp1} and going to the Coulomb vacuum $|\text
C\rangle = |N_{I}=1;\nu_{q}=0\rangle$. Symmetries are no longer enough
when $\Nf=2N$ to fix completely the solution to the anomaly equations
in this vacuum, but we can rely on explicit instanton calculations
\cite{nekrasov}. Introducing the polynomial
\be\label{Udef} U(z) =\prod_{q=1}^{2N}(z-m_{q})\, ,\ee
it can be shown that the characteristic function \eqref{defF}
satisfies
\be\label{eqF2}\langle\text C|\F(z)|\text C\rangle +
\frac{h\, U(z)}{\langle\text C|\F(z)|\text C\rangle} = P_{|\text
C\rangle}(z)\ee
for some degree $N$ polynomial $P_{|\text C\rangle}$. The conditions
obtained by expanding at large $z$ and writing that the negative
powers in $z$ in the left hand side of \eqref{eqF2} must vanish
determine recursively the polynomials $\mathscr P_{p}$. Conversely,
the operator relations \eqref{qrel} then implies that
\be\label{eqF3}\langle\F(z)\rangle + \frac{h\,U(z)}{\langle\F(z)\rangle}
= P(z)\ee
in all the vacua, for a certain vacuum-dependent polynomial $P$. This
in turn yields the quantization conditions \eqref{qc1} and \eqref{qc2}
(or the appropriate factorization of the associated Seiberg-Witten
curves) in full generality.
\section{Conclusion and open problems}
\setcounter{equation}{0}

In this paper, we have obtained an extremely simple interpretation of
the quantization conditions \eqref{qc1} and \eqref{qc2}. These
conditions simply encode the precise form of vacuum-independent
operator relations between the chiral observables $u_{k}=\tr X^{k}$.
Since the relations are completely fixed by studying a weakly coupled
Coulomb vacuum, the validity of \eqref{qc1} and \eqref{qc2} in all the
vacua of the $\nn=1$ theory, including the strongly coupled confining
vacua, can be derived from a purely semi-classical analysis. It is
particularly startling that such a conceptually simple understanding
can be achieved. It yields in particular a straightforward
justification from first principles of the well-known strong coupling
approach.

A natural question is whether the same ideas can be used to derive the
quantization conditions in the Coulomb vacuum as well, in the most
general cases, and independently of the semi-classical approximation.
This seems plausible, because it is absolutely not obvious a priori
that the solutions to the anomaly equations can be consistent with the
existence of \emph{vacuum-independent} relations between the
variables. This consistency requirement does not arise in the
similar-looking loop equations of the planar matrix model, because in
this case all the variables are independent. In the gauge theory, it
is natural to conjecture that consistency can be achieved if and only
if the conditions \eqref{qc1} and \eqref{qc2} are satisfied in the
Coulomb vacuum (and thus in all the other vacua by the arguments
above).

An outstanding open problem is to provide a non-perturbative proof of
the generalized Konishi anomalies. The existing derivations are made
in perturbation theory, with a fixed classical background gauge field
\cite{CDSW}. Contrary to a statement often made in the literature, the
equations do get non-perturbative corrections, but it is believed that
these corrections can be made implicit by suitably defining the
variables, as explained in Section 2. Since the equations must be
valid in all the vacua, it is enough to make a proof in the context of
instanton calculus, and this is presently under investigation.

In \cite{fer0}, it was also shown, from another point of view, that
the quantization conditions \eqref{qc1} and \eqref{qc2} are not
dynamical but rather follow from general consistency conditions. In
this respect, \cite{fer0} and the present work share the same
philosophy. The argument of \cite{fer0} was based on the analysis of
the gauge invariance of the operator $\F(z)=\det(z-X)$ for all values
of $z$. Gauge invariance turns out to be consistent with the analytic
continuation in $z$ if and only if the conditions \eqref{qc1} and
\eqref{qc2} are satisfied. However, gauge invariance of $\det (z-X)$
can be achieved only for some particular definitions of the variables
$u_{N+p}$, and thus is not trivial a priori (this is similar to the
fact that the quantization conditions \eqref{qc2} are not trivial a
priori). Again, an explicit non-perturbative definition of $\det(z-X)$
can certainly be given in the context of the instanton calculus. As
suggested in \cite{fer0}, the standard representation of the
determinant in terms of a fermionic integral is then likely to make
gauge invariance manifest. Combined with the non-perturbative analysis
of the anomaly equations, the arguments of \cite{fer0} would then
provide a new and elegant way to sum up instanton series explicitly,
independently of the localization methods used in \cite{nekrasov}.

\subsection*{Acknowledgements}

This work is supported in part by the belgian Fonds de la Recherche
Fondamentale Collective (grant 2.4655.07), the belgian Institut
Interuniversitaire des Sciences Nucl\'eaires (grant 4.4505.86),
the Interuniversity Attraction Poles Programme (Belgian Science
Policy) and by the European Commission FP6 programme
MRTN-CT-2004-005104 (in association with V.\ U.\ Brussels). The author
is on leave of absence from Centre National de la Recherche
Scientifique, Laboratoire de Physique Th\'eorique de l'\'Ecole Normale
Sup\'erieure, Paris, France.

\renewcommand{\thesection}{\Alph{section}}
\renewcommand{\thesubsection}{\arabic{subsection}}
\renewcommand{\theequation}{A.\arabic{equation}}
\setcounter{section}{0}
\end{document}